\documentclass[emulateapj]{emulateapj}

\begin{document}
\title{Direct Detection and Orbit Analysis of the Exoplanets HR 8799 \lowercase{bcd} \\from Archival 2005 Keck/NIRC2 Data}
\author{Thayne Currie\altaffilmark{1}, Misato Fukagawa\altaffilmark{2},  
Christian Thalmann\altaffilmark{3}, Soko Matsumura\altaffilmark{4}, 
Peter Plavchan\altaffilmark{5}} 
\altaffiltext{1}{NASA-Goddard Space Flight Center}
\altaffiltext{2}{Osaka University}
\altaffiltext{3}{Astronomical Institute Anton Pannekoek, University of Amsterdam}
\altaffiltext{4}{Department of Astronomy, University of Maryland-College Park}
\altaffiltext{5}{NExScI, California Institute of Technology}
\begin{abstract}
We present previously unpublished July 2005 $H$-band coronagraphic data of the young, planet-hosting star HR 8799 
from the newly-released Keck/NIRC2 archive.  Despite poor observing conditions, we 
detect three of the planets (HR 8799 bcd), two of them (HR 8799 bc) without advanced image processing.  
Comparing these data with previously published 1998-2011 astrometry and that from 
re-reduced October 2010 Keck data constrains the orbits of the planets.  
Analyzing the planets' astrometry separately, HR 8799 d's orbit is likely inclined at least 25$^\circ$ from face-on and the 
others may be on in inclined orbits.  For semimajor axis ratios consistent with a 
4:2:1 mean-motion resonance, our analysis yields precise values for HR 8799 bcd's orbital parameters and strictly 
constrains the planets' eccentricities to be less than 0.18--0.3.
  However, we find no acceptable orbital solutions with this resonance that place the planets 
in face-on orbits; HR 8799 d shows the largest deviation from such orbits.
Moreover, few orbits make HR 8799 d coplanar with b and c, whereas dynamical stability analyses used to constrain the planets'
masses typically assume coplanar and/or face-on orbits. 
  This paper illustrates the significant science gain enabled with the release of the NIRC2 archive.
\end{abstract}
\keywords{planetary systems, stars: early-type, stars: individual: HR 8799} 
\section{Introduction}
The nearby, young A-type star HR 8799 \citep[$d$ = 39.4 $pc$, $\approx$ 30 Myr;][]{Zuckerman2011} harbors the first independently confirmed, directly imaged
 exoplanetary system and the only imaged multi-planet system \citep{Marois2008}.  After the discovery 
of HR 8799 bcd ($r_\mathrm{proj}$ $\approx$ 24, 38, and 68 AU) reported in November 2008 \citep{Marois2008}, 
other studies identified at least one of these planets 
in archival data taken prior to 2008 \citep{Lafreniere2009,Fukagawa2009,Metchev2009,Soummer2011}.  

HR 8799 planet astrometry derived from both pre and post-discovery images 
can help constrain the system's dynamical 
stability and, in turn, the planets' physical properties.  At least two of the HR 8799 planets 
are likely locked in a mean motion resonance, otherwise the system would quickly become dynamically unstable 
\citep{Fabrycky2010}.  The recently discovered fourth companion at $\sim$ 15 AU, HR 8799 e,
generally makes dynamical stability less likely \citep{Marois2011,Currie2011a}, favoring lower masses 
of $M_\mathrm{b,cde}$ $<$ 7, 10 $M_\mathrm{J}$,
an important constraint given the uncertainties in deriving masses from planet cooling and atmosphere models
\citep[][]{Spiegel2012,Madhusudhan2011}.  

Studies focused on fitting the planets' orbits and/or testing dynamical stability typically assume that the planets are 
a) in resonance (4:2:1 for HR 8799 bcd or 2:1 for HR 8799 cd), b) in circular, face-on orbits, c) and/or in coplanar orbits 
\citep[e.g.][see also Fabrycky and Murray-Clay 2010]{Marois2011,Currie2011a}.  However, \citet{Soummer2011} show that circular, face-on, and 
coplanar orbits are inconsistent with 1998 HST astrometry, identifying a best-fit orbit for HR 8799 d of $i$ = 28$^\circ$ and 
$e$ = 0.115.  Generally, more eccentric orbits destabilize the system.  The system stability depends on the (mutual) inclinations 
of the planets \citep{Fabrycky2010,Sudol2012}.  Thus, the HR 8799 planets' true mass limits derived from 
dynamical stability arguments may slightly differ from those previously reported.  
 
Well-sampled HR 8799 d astrometry could help clarify whether HR 8799 d's orbit must be 
inclined, eccentric, and/or coplanar with the other planets.  However, until now there 
is a $\sim$ 9-year gap between the 1998 HST detection and the next one (2007; Metchev et al. 2009).  New astrometry 
for HR 8799 bce in between 1998 and 2007 could also help constrain those planets' orbits.
  By better determining the HR 8799 planets' orbital properties,
 we can more conclusively investigate system dynamical stability and thus
 better clarify the range of allowable planet masses.

In this Letter, we report the detection of HR 8799 bcd from unpublished, now-public 
Keck/NIRC2 data taken in 2005 supplemented with a re-reduction of published October 2010 data from \citet{Marois2011}.  We use 
these data to better constrain the orbital properties of HR 8799 bcd.

\section{Observations and Data Reduction}
\subsection{July 2005 Data}
We downloaded HR 8799 data taken on July 15, 2005 from the newly-available Keck/NIRC2 data archive 
(Program ID H53BN2, P.I. Michael Liu).  The data were taken in $H$ band with the narrow camera 
\citep[9.952 mas/pixel;][]{Yelda2010} with the 0\farcs6 diameter coronagraphic spot and the 
``incircle" pupil plane mask.  HR 8799 was observed in 10\,s exposures in ``vertical angle" or \textit{angular differential 
imaging} mode \citep{Marois2006} through transit (Hour Angle = [$-$0.20,0.55]) with a total field rotation of 
147.1$^{\circ}$.  During these observations, the seeing conditions fluctuated and the observers 
periodically recentered the star behind the mask, changing the intensity profile of the stellar halo 
(and thus the quasi-static speckle pattern).   During a few frames near transit the star 
did not properly center at all behind the coronagraph.  We identify $\sim$ 11 minutes of 
science-grade data.  Basic image processing followed standard steps previously 
used to process NIRC2 data \citep{CurrieRodigas2012}.

\begin{figure}
\centering
\includegraphics[scale=0.36,trim=0mm 0mm 15mm 2mm, clip]{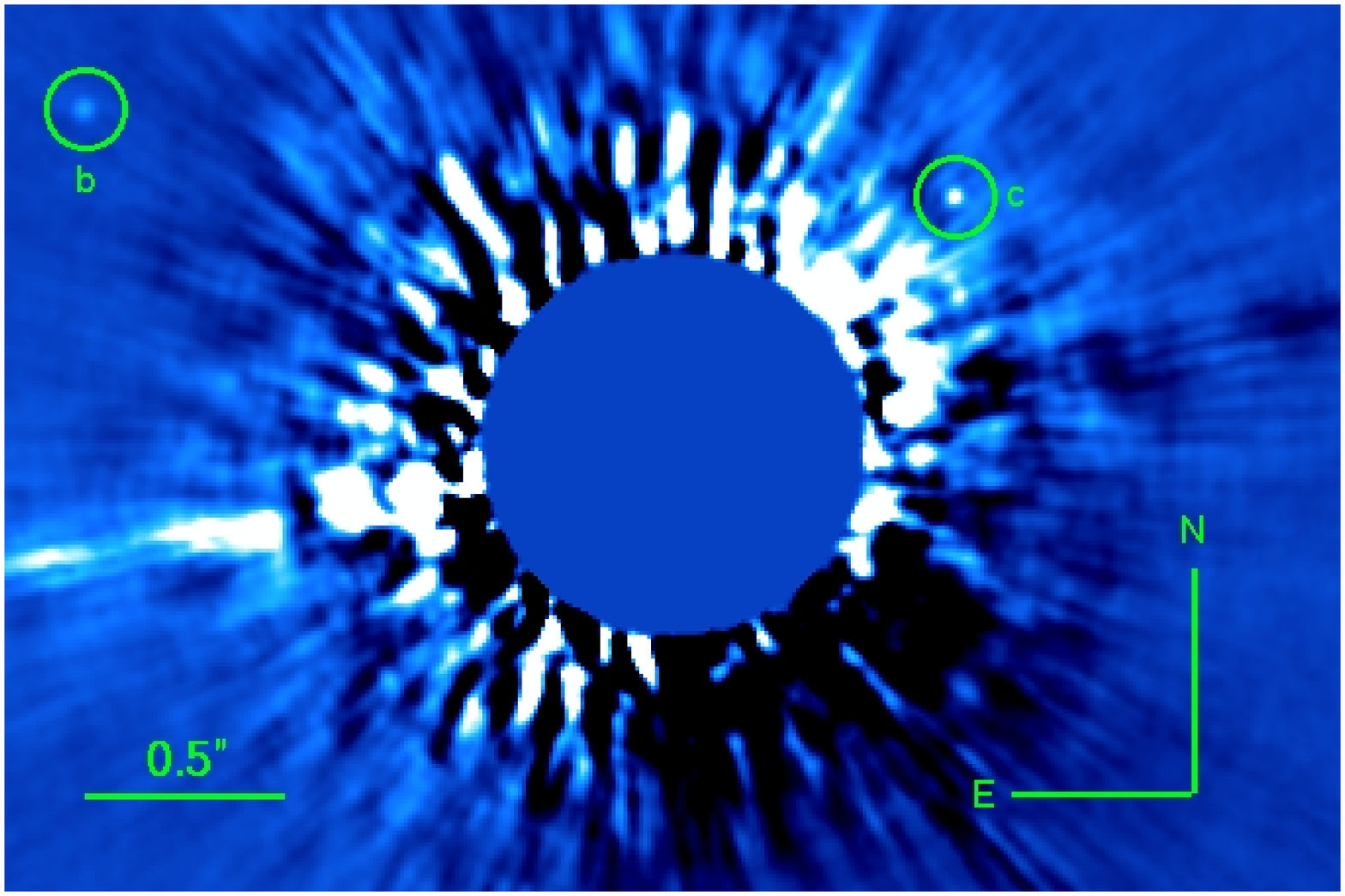}
\includegraphics[scale=0.36,trim=0mm 0mm 15mm 2mm,clip]{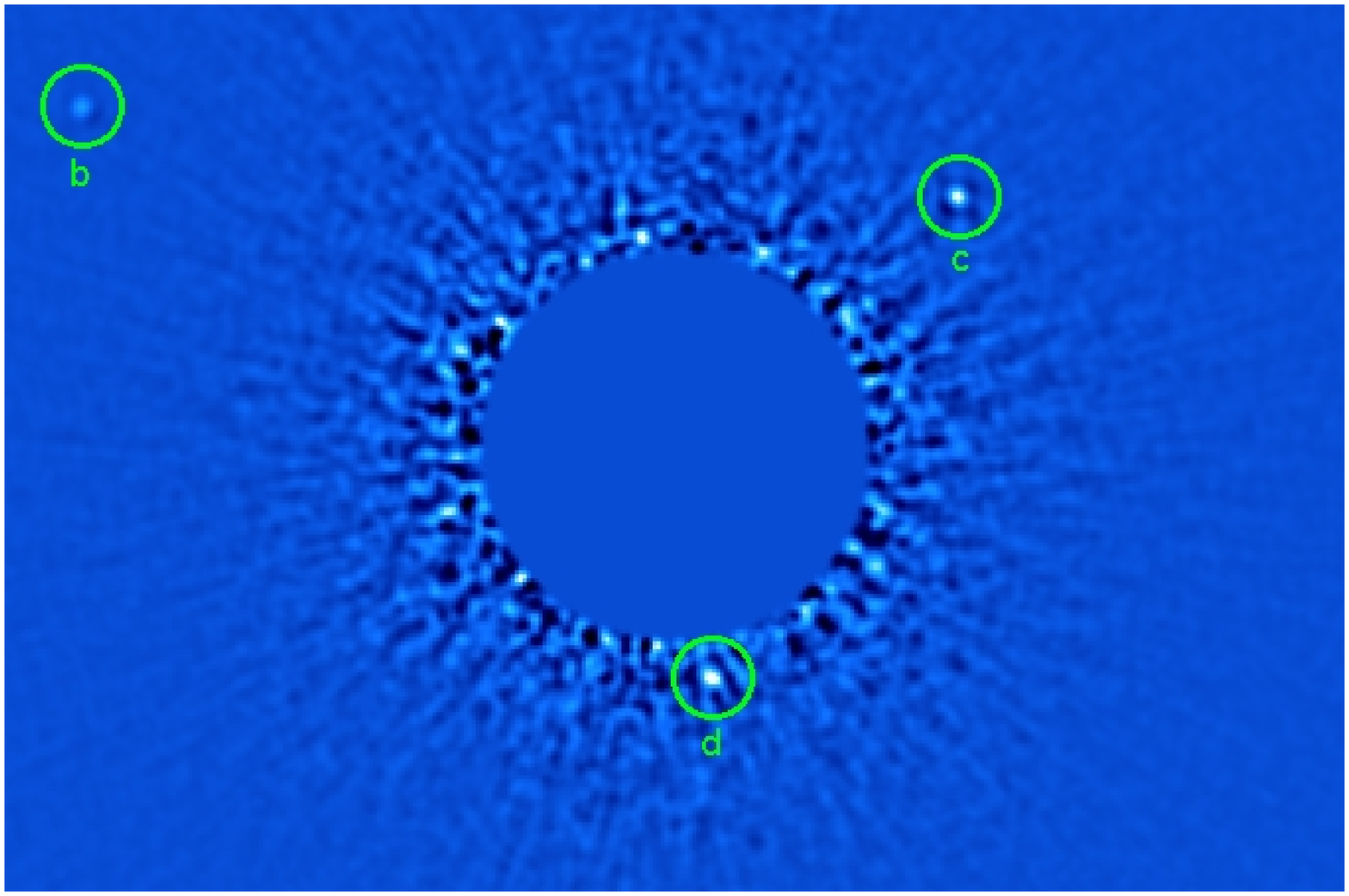}
\caption{ HR 8799 images processed with ``classical" PSF subtraction (left) and 
A-LOCI (right) showing the detections of HR 8799 $b$, $c$, and $d$ (circled).}
\label{keckimages}
\end{figure}

For a first-order reduction, we perform a simple
 ``classical" ADI-based PSF subtraction \citep[e.g.][]{Marois2006}.
Figure \ref{keckimages} shows this reduction, clearly revealing HR 8799 b (SNR $\sim$ 12) and 
identifying HR 8799 c, albeit at low SNR ($\sim$ 4).  
With the LOCI approach \citep{Lafreniere2007} as 
implemented and modified in previous work \citep{Currie2010,Currie2011a,Currie2011b}, 
we easily detect HR 8799 b and c and obtain a marginal detection of HR 8799 d at $r$ $\sim$ 0\farcs6 
several degrees away (clockwise) from the July 2007 position reported by \citet{Metchev2009} (not shown).

To improve the signal-to-noise of our HR 8799 d detection, we incorporated several 
 upgrades to enhance contrast which are being implemented in a new ``adaptive" LOCI (A-LOCI) 
pipeline (T. Currie, 2012 in prep.; see also similar steps in Marois et al. 2010b).
We subtract off the seeing halo in each image to measure the 
static/quasi-static speckle pattern, determining the cross-correlation function for the speckle patterns 
in annular sections for all possible image pairs.  While the speckle pattern is generally better correlated between 
frames taken close together in time, this is not always the case, especially when comparing frames before and after 
small telescope nods.  Therefore, for each annular section of a science image we want to subtract, we filter 
reference image sections by their degree of correlation to remove those below a certain, predetermined threshold ($r_\mathrm{corr}$).  

HR 8799 b is detectable in most individual processed frames (SNR/frame $\sim$ 4--7), so 
we measure its position to identify and correct for any astrometric biases caused by 
a PA ``jump" for frames obtained near 
transit due to imperfect mechanical alignment of the telescope's $y$ axis.
The position angles of HR 8799 b in frames more than 0.25 hours from transit are consistent, but 
the PA offset follows a bell-shaped curve with a maximum offset of $\approx$ 0.6 degrees centered on transit.  
We model and correct for this offset using a fifth-order polynomial.  We also reran our pipeline with 
different rotation axis offsets due to image registration errors, setting an upper limit 
to this of 0.5 pixels in each coordinate.

Furthermore, unlike the original LOCI algorithm, we set the azimuthal length of the subtraction zone to be smaller (not identical) to the 
azimuthal length for the optimization zone, equal to $dr$ (the length along the radial direction).  We then center the optimization zone 
on the subtraction zone.  Finally, we iteratively determine the algorithm parameters -- $\delta$, $N_{A}$, 
$g$, $dr$ \citep[][]{Lafreniere2007}, and $r_{corr}$ -- 
that maximize the signal-to-noise of point sources and applied these settings to extract our final image\footnote{We also considered a 
  ``reference PSF library" derived from other 2004--2005 NIRC2 data obtained with the same setup 
(coronagraph size, filter, etc.) to further attenuate speckles.   However, this library degrades the SNR 
of HR 8799 b and c by $\approx$ 30--60\% and renders HR 8799 d undetectable, because HR 8799's speckle patterns are poorly correlated with 
library's.}.  

Figure \ref{keckimages} (right panel) shows our final image displaying higher signal-to-noise detections of HR 8799 b (SNR $\sim$ 38) and 
c (SNR $\sim$ 18) and new detection of d (SNR $\sim$ 5), using algorithm parameters of
$\delta$ $\ge$ 0.74, $N_{A}$ = 245, $g$ = 0.95, $dr$ = 6, and $r_{corr}$ $\ge$ 0.315, though 
similar settings yield nearly identical results (i.e. $\delta$ $\ge$ 0.73--0.8, $r_{corr}$ $\ge$ 0.31--0.36).
We achieve contrast gains of up to $\sim$ 80\% (for HR 8799 d) 
over our best LOCI reduction.  

For flux calibration, we perform aperture photometry on HR 8799 bcd and on the nearby star GJ 616.2 observed just prior 
to HR 8799.  We use fake point sources to further correct for LOCI-based photometric biases \citep[i.e.][]{Lafreniere2007,
Currie2011a,Currie2011b} and find $m(H)$ = 18.05 $\pm$ 0.09 mag for HR 8799 b, $m(H)$ = 17.06 $\pm$ 0.13 for HR 8799 c, and $m(H)$ = 
16.71 $\pm$ 0.24 for HR 8799 d.  The magnitude differences between HR 8799 c and 
d appear slightly discrepant compared to better calibrated NIRC2 measurements from \citet[][]{Marois2008}, 
 although the individual measurements are consistent to within $\sim$ 1\,$\sigma$.  
Photometry derived for HR 8799 bc using classical PSF subtraction agrees with that derived from our A-LOCI based reduction 
within errors.  
Bright residual speckles at $r$ = 0\farcs3--0\farcs4 prevent detecting HR 8799 e.

\subsection{October 2010 Data}
To supplement the 2005 HR 8799 astrometry, we downloaded and reduced October 2010 $L^\prime$-band NIRC2 
data from the Keck archive (P.I. B. Macintosh).  These data are the latest reported by \citet{Marois2011} who focus on HR 8799 e astrometry: we use 
these data instead to extract astrometry for HR 8799 bcd.  Individual exposures consist of 50\,s frames totaling $\approx$ 80 minutes 
taken through transit without a coronagraph with periodic telescope nods for sky subtraction.  
  Observing conditions appeared variable at a level comparable 
to the July 2005 data and worse than other recent Keck data 
\citep[i.e.][]{CurrieRodigas2012,Rodigas2012}, but we detect all four planets (SNR $\sim$ 6--20) with our pipeline.  

\section{Astrometric Analysis and Orbit Fitting}
\subsection{Method}
We use our detections to better constrain the orbits of HR 8799 bcd, first by fitting the orbits of 
the planets separately, then identifying the subset of orbits consistent with systematically more stable 
mean motion resonance configuration.  We calibrate our astrometry 
by assessing and correcting for biases introduced by LOCI-based processing in the same manner as our 
photometric calibration, using fake point sources.  
Using different telescopes and slightly different image processing techniques lead to systematic biases in planet astrometry.  To minimize these biases, 
we restrict ourselves to astrometry from HST/1998 \citep{Soummer2011}, Subaru/2002 and 2009 \citep{Fukagawa2009, Currie2011a}, Keck/2004, 2005, 2007--2009 
\citep[][ this work]{Marois2008, Metchev2009,Galicher2011}, VLT/2009 \citep{Currie2011a}, and LBT/Pisces \citep{Esposito2012}.
For the Pisces data, we include the substantial north position angle uncertainty.
  We furthermore 
modify the Keck/NIRC2 astrometry reported in \citet{Marois2008} and \citet{Metchev2009} to reflect the updated NIRC2 astrometric 
calibration \citep{Yelda2010} 
accuracy, rescaling the position by a 
factor of 9.952/9.963, and putting in a $PA_\mathrm{north,new}$-$PA_\mathrm{north,old}$ = 0.13$^{\circ}$ clockwise rotation.

To separately determine the range of allowable HR 8799 bcd orbits, we follow our previously-adopted Monte Carlo-based 
approach \citep[][]{Thalmann2009,Currie2011b}, comparing the HR 8799 planet astrometry to 
predictions from randomly-selected orbits.  In a first set of ``conservative" simulations, we consider the 
orbits separately.  Dynamical stability analysis suggests that HR 8799 bcd are likely in resonance 
\citep{Fabrycky2010}; the 4:2:1 resonance is particularly adept at stabilizing the system.
  To focus on dynamically stable orbits we then select the subset of 
best-fitting orbits that preclude the planets from crossing orbits and are consistent with a 4:2:1 resonance.  
Here we define ``resonance" broadly, including orbits with ratios of periods between 1.9 and 2.1 for consecutive pairs of planets since, 
at least in some circumstances, \textit{exact} period ratios may be rare \citep[e.g.][]{Fabrycky2012}
 \footnote{Dynamical simulations identify stable solutions for at least \textit{some} planet masses where only HR 8799 c and d are 
in resonance, so our conclusions from this set of astrometric analyses may be less applicable for HR 8799 b.}.

For all our simulations, the minimum $\chi^{2}$ for the b, c, and d planets are 20.8, 13.8, and 14.8, for 
reduced $\chi_\nu^2$ values of 1.04, 0.86, and 1.24.  
Following \citet{Currie2011b}, we choose a cutoff of $\chi^{2}$ $\le$ $\chi_\mathrm{min}^{2}$ + 1 to represent the family 
of best-fitting orbits.  Formally, this cutoff admits only an average additional deviation per each x or y measurement of 
$\approx$ 1/2$N_\mathrm{obs}$ (i.e. $\approx$ 1/20 for HR 8799 b; $\approx$ 1/14 for HR 8799 d) beyond the best-fit 
models which themselves imply typical deviations of $\approx$ 1-$\sigma$ 
per each x or y measurement (since the minimum reduced $\chi_\nu^{2}$ values are $\approx$ 1).  However, we obtain 
nearly identical results for more relaxed cutoffs (see below).  
From the set of models passing our $\chi^{2}$ cutoff, including the subset in resonance, we 
determine the weighted median value and the weighted 68\% confidence interval about the median for each model parameter 
from amongst the set of best-fitting orbits.

Furthermore, we report a ``\emph{most likely orbit}" (MLO) simply defined as follows. First, over the best-fitting 
family of orbits, we calculate histograms for the following parameters: 
logarithm of the semi-major 
axis ($h_{\log a}$), eccentricity ($h_e$), inclination ($h_i$), longitude of ascending node 
($h_\Omega$), and argument of periastron ($h_\omega$). For each orbit $n$ in the best-fitting 
ensemble, we then define the measure of likelihood $\mathcal{L}(n)$ as
\begin{equation}
\mathcal{L}(n) = h_{\log a}(\log a_n) \cdot h_e(e_n) \cdot h_i(i_n) \cdot h_\Omega(\Omega_n) \cdot h_\omega(\omega_n) \cdot W,
\end{equation}
\noindent i.e.,\ the product of all histograms values in the bins in which the orbit $n$ lies, 
representing the individual likelihood of each measured orbital parameter, as well as the 
statistical weight, $W$, representing the likelihood of the observed planet position within 
the orbit (i.e.,\ the anomaly)\footnote{Here, the statistical weight $W$ 
is defined as the mean orbital velocity for the corresponding 
orbit divided by the orbital velocity at the observed epoch, 
$W := \langle{}v\rangle{}_\mathrm{orbit} / v(t_\mathrm{obs})$.}.
The most likely orbit is then the one orbit that maximizes 
the measure of likelihood, $\mathcal{L}(n_\mathrm{MLO}) = \max_n \mathcal{L}(n)$. Because of
 the highly skewed distribution of some parameters from the best-fit orbits, 
in particular $\log a$, the MLO parameters can differ significantly from the weighted median parameters.


\subsection{Results}
\begin{figure}
\centering
\includegraphics[scale=0.7,clip]{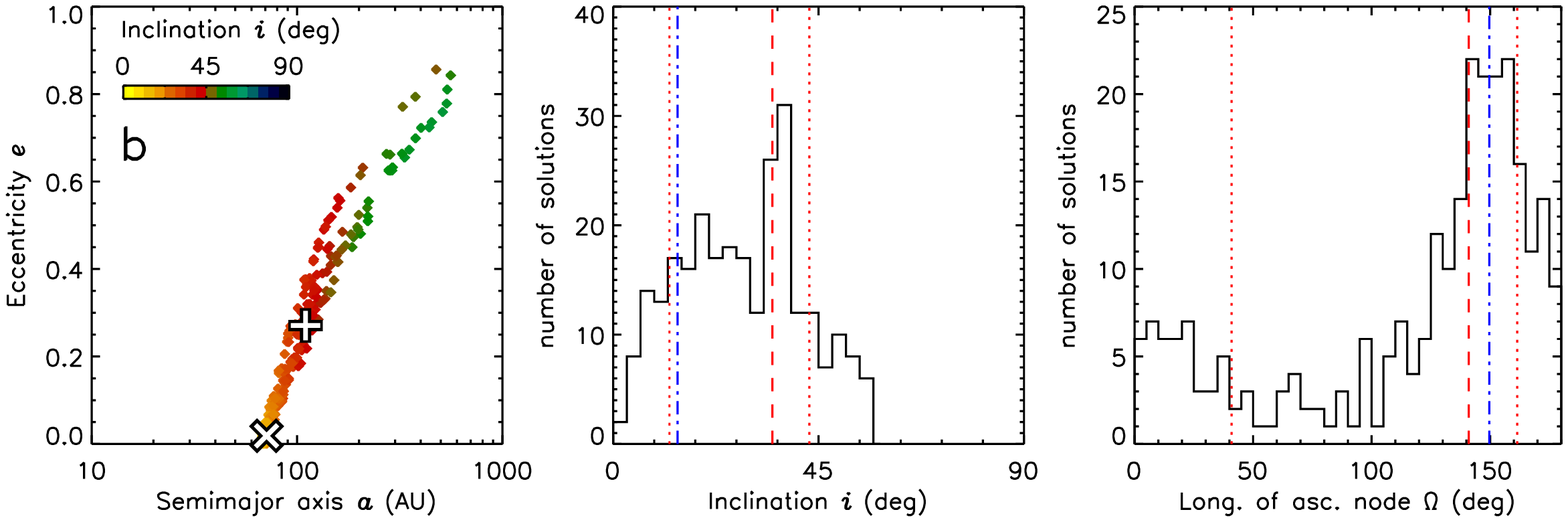}
\vspace{-0.05in}
\\
\includegraphics[scale=0.7,clip]{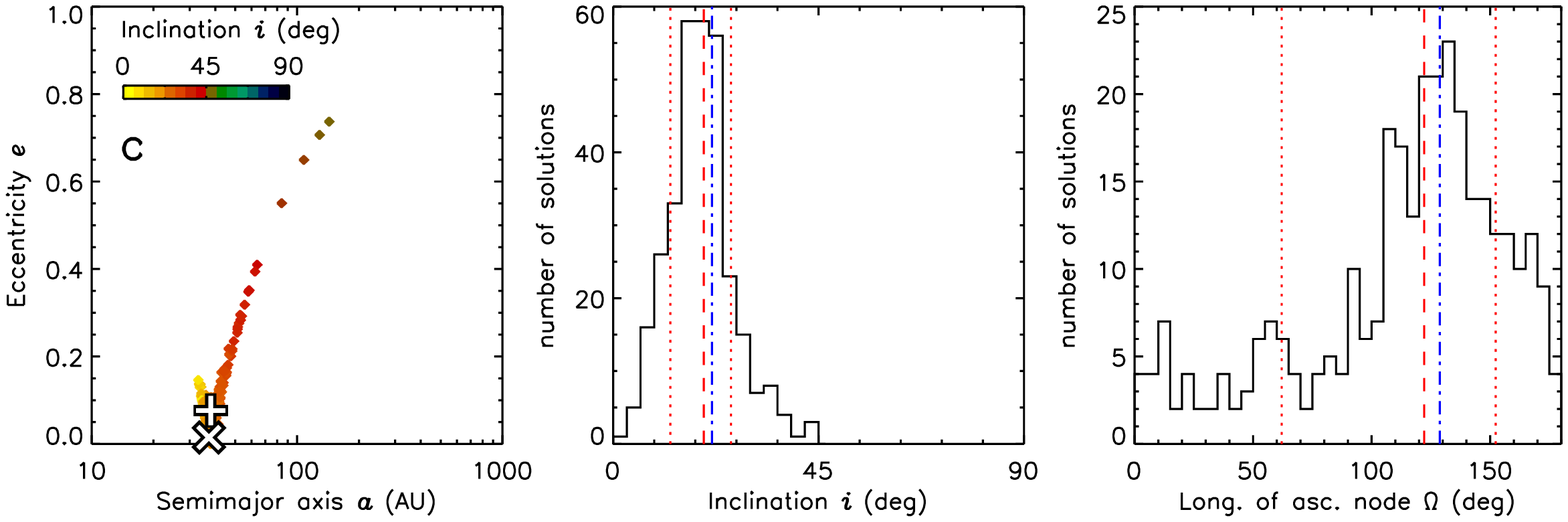}
\vspace{-0.05in}
\\
\includegraphics[scale=0.7,clip]{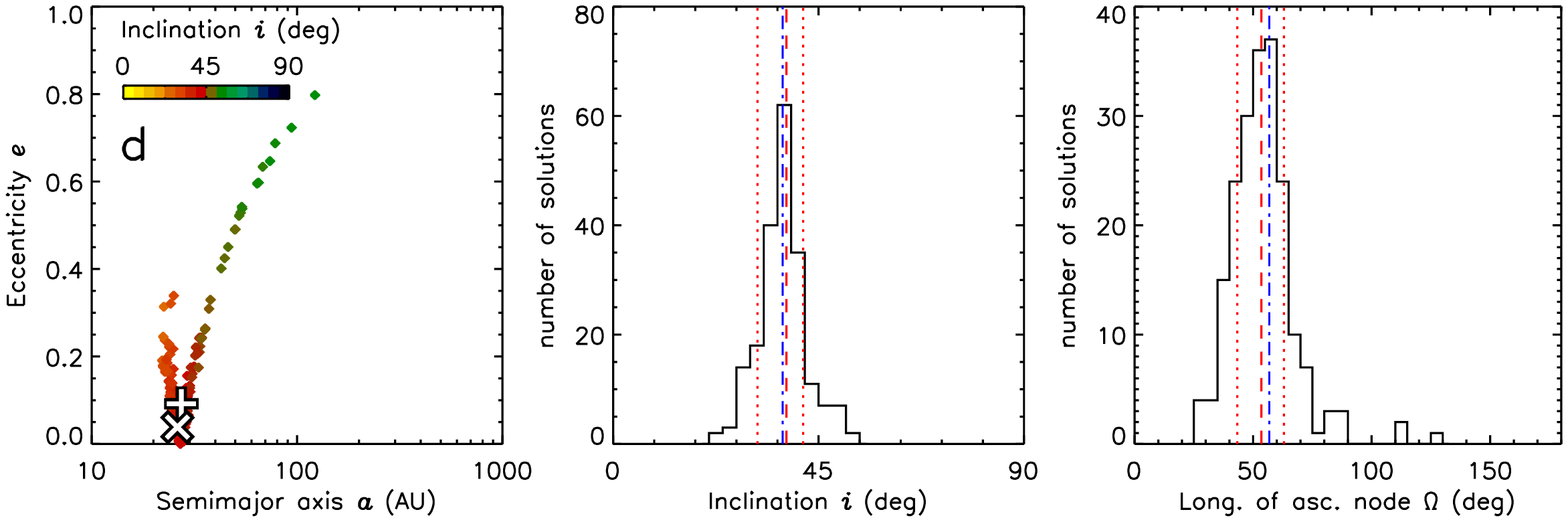}

\vspace{-0.1in}
\caption{Astrometric analysis results for HR 8799 b (top), HR 8799 c (middle), and HR 8799 d (bottom)
 considering the full range of orbits satisfying the criterion $\chi^{2}_{\nu}$ $<$ 
$\chi^{2}_{\nu, min}$ +1.  The panels show the orbits in $a$/$e$/$i$ space (left) and histogram 
distributions of the orbital inclination $i$ (middle panels) and longitude of ascending node (right panels).  
In the left panels, the `x' identifies parameters $e$ and $a_{p}$ from the best-fit orbital solution; the `plus' sign denotes 
the weighted median value for the same parameters.  For the middle and right panels, 
the vertical blue dashed line identifies the MLO, the vertical red dashed line identifies the 
median parameter value, and the vertical red dotted lines bracket the 68\% confidence interval.}
\label{astrom_all}
\end{figure}

Table 1 summarizes our results, and Figure \ref{astrom_all} displays the orbits in $a$/$i$/$e$ 
space (left), the histogram distribution of $i$ (middle), and the histogram distribution of the longitude of the 
ascending node, $\Omega$ (right).  The top panels display properties for HR 8799 b, the middle for HR 8799 c, and 
the bottom for HR 8799 d.  Assuming a $\chi^{2}$ cutoff of $\chi_\mathrm{min}^{2}$ + 1, 
the observations well constrain the HR 8799 d orbit --
 $a$ $\sim$ 24--32 AU, $i$ $\sim$ 32--42$^{\circ}$, $e$ $\sim$ 0.03--0.23, and $\Omega$ $\sim$ 43--63$^{\circ}$.  
They limit the HR 8799 c's most plausible orbital parameters to $a$ $\sim$ 36--42 AU, $i$ $\sim$ 13--26$^{\circ}$, and $e$ $\sim$ 0.03--0.13.
As expected, the parameters for HR 8799 b are the most poorly constrained, showing the widest dispersion and the largest 
differences between the median parameter value and that from the MLO.

Our analysis clearly disfavors face-on orbits for all three planets, especially for HR 8799 c and d.
Additionally, the inclination distributions for HR 8799 bc appear systematically skewed towards values lower than those for d.
Formally, though, the set of acceptably-fitting orbits HR 8799 bcd include some that make
 the planets coplanar.  

\begin{figure}
\centering
\includegraphics[scale=0.7,clip]{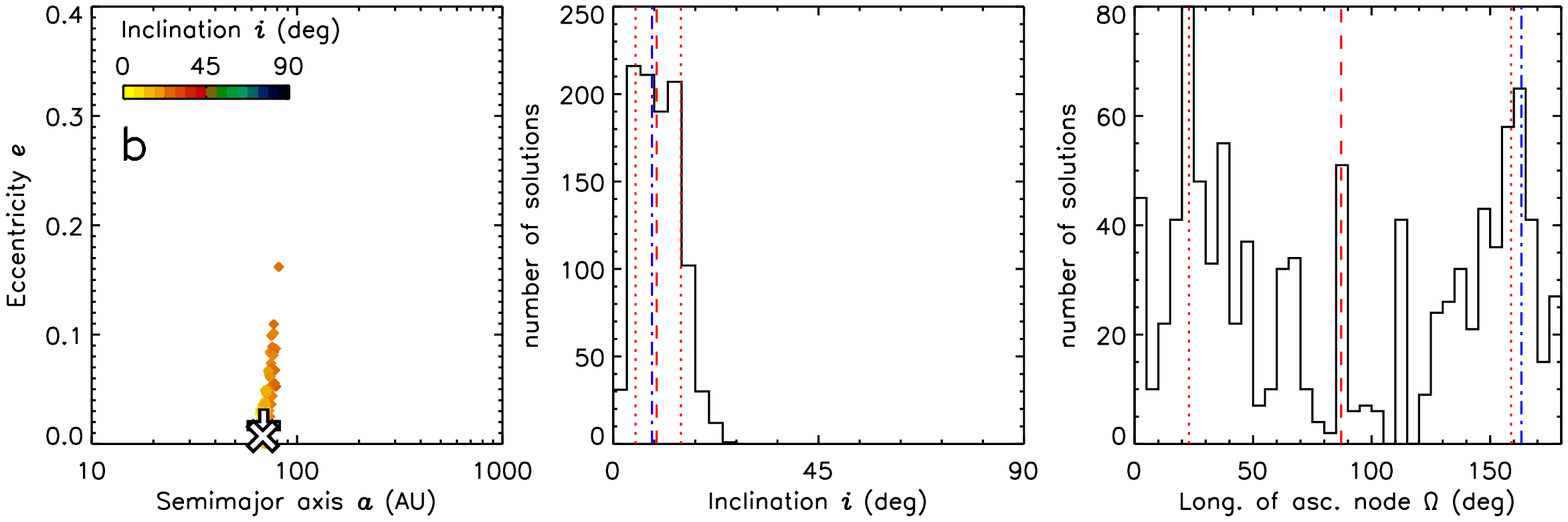}
\vspace{-0.05in}
\\
\includegraphics[scale=0.7,clip]{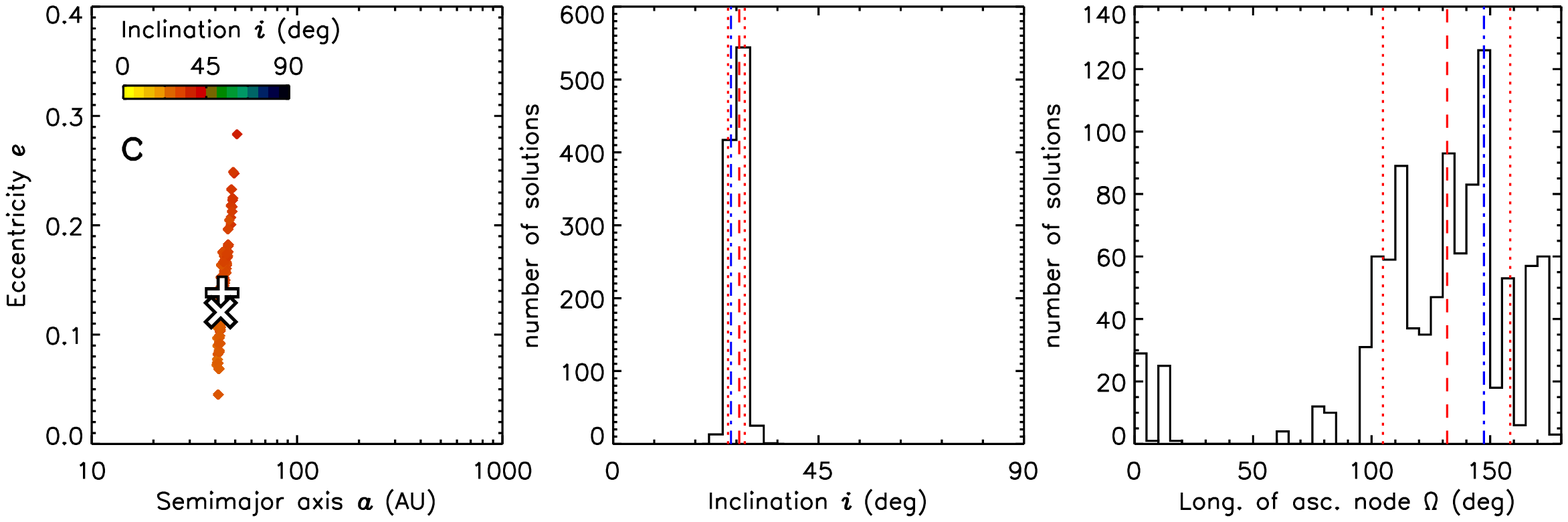}
\vspace{-0.05in}
\\
\includegraphics[scale=0.7,clip]{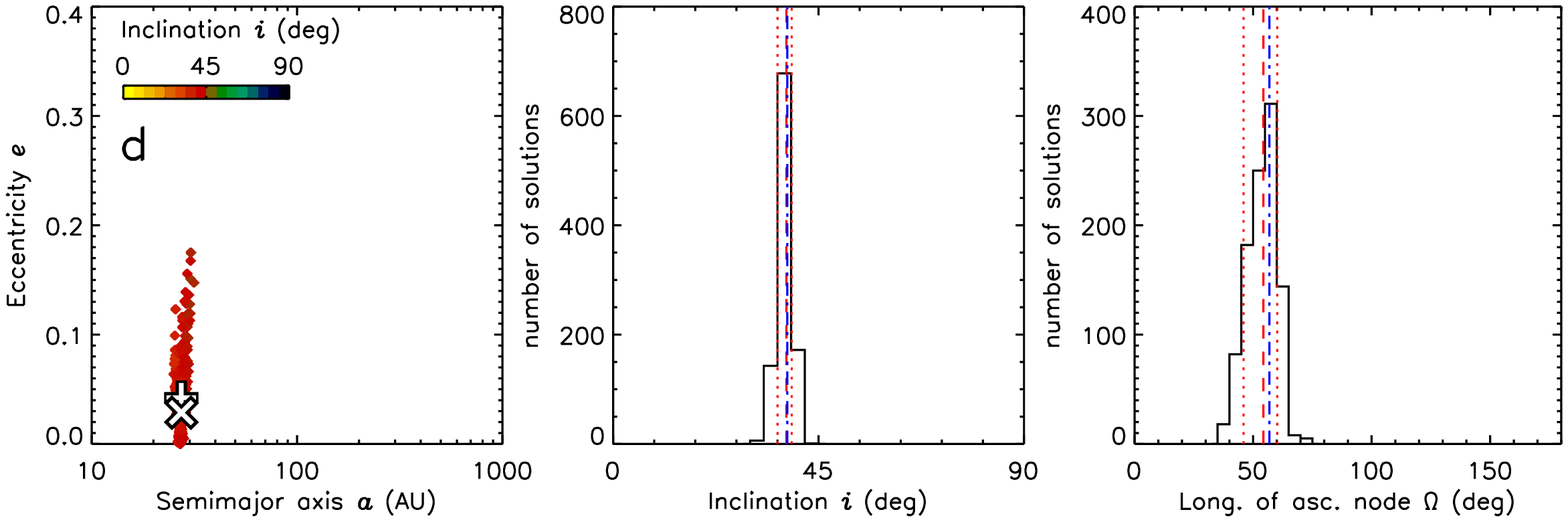}
\vspace{-0.1in}
\caption{Same as Figure \ref{astrom_all} except only for non-crossing orbits consistent with a 4:2:1 mean-motion resonance between 
HR 8799 b, c, and d, a configuration which promotes orbital stability \citep[e.g][]{Fabrycky2010}.  Note the lack of 
high-eccentricity orbits and the narrower range in acceptable orbital parameters, especially for HR 8799 c and d.}
\label{astrom_res}
\end{figure}

Assuming a 4:2:1 mean-motion resonance configuration, we place far stronger limits on nearly all of the HR 8799 bcd orbital 
properties (Figure \ref{astrom_res}).  This assumption explicitly rules out $e$ $>$ 0.18 for HR 8799 bc 
and $e$ $>$ 0.3 for HR 8799 d.  Likewise, we identify a very narrow range of planet semimajor axes: $a$ = 
67.5--70.8 AU, 42.1--44.4 AU, and 26.4--28.1 AU for HR 8799 b, c, and d.  The 68\% confidence interval in $\Omega$ for 
HR 8799 d further narrows to 46$^\circ$--62$^\circ$.

For this configuration, HR 8799 d (c) must be in an orbit viewed more than 25$^\circ$ 
(15$^\circ$) from face-on while 
HR 8799 b is likely inclined by at least 5$^\circ$.  Furthermore, the inclination distributions between 
HR 8799 d and HR 8799 bc are even more dissimilar, implying that HR 8799 d is most likely inclined relative to 
c by at least $\approx$ 7$^\circ$ and b by more than $\approx$ 21$^\circ$.  While our analyses cannot conclusively 
rule out coplanar orbits in a 4:2:1 mean-motion resonance, they suggest that few such orbits 
are compatible with 12 years of HR 8799 planet astrometry.

To confirm that we are fully sampling the subset of orbital parameters covering the $\chi^{2}$ minima, we run 
our simulations with a more relaxed $\chi^{2}$ cutoff of $\chi^{2}$ $\le$ $\chi_\mathrm{min}^{2}$ + [5, 5, 3.5] for 
HR 8799 b, c, and d, which formally admit an additional average deviation from the data of 
$\approx$ 0.3-$\sigma$ for each measurement.  With this cutoff, we obtain nearly identical results (second set of 
rows in Table 1).  Considering the planets' orbits separately, we find $a$ $\sim$ 23--31 AU, $i$ $\sim$ 27--41$^\circ$, 
$e$ $\sim$ 0.01--0.31, and $\Omega$ $\sim$ 41--53$^\circ$ for HR 8799d.  We 
find similar ranges in orbital parameters for HR 8799 c and (for the resonance case) 
HR 8799 b.  Likewise, the planets' range of inclinations exclude face-on orbits.  
HR 8799 d's inclination distribution is skewed to systematically higher values, expected if it is 
non-coplanar with b and c, although here there are more orbit combinations that could make the planets coplanar.  

\section{Discussion}
From analyzing HR 8799 bcd astrometry from our new ``pre-discovery" image and other data, we 
provide new constraints on the planets' orbital properties.
Treating the three planets separately, we narrowly constrain three major orbital parameters 
($a$/$i$/$e$) for HR 8799 c and d.  None of the planets are likely to be orbiting 
face-on and the inclinations for acceptably-fitting orbits are systematically higher for HR 8799 d than for 
HR 8799 b and c.  

If HR 8799 bcd have semimajor axes consistent with a 4:2:1 resonance, our analysis strongly constrains the major orbital properties 
for all three planets.  The three planets (especially c and d) then even more obviously have inclined orbits.
Most acceptable solutions for HR 8799 d place the planet on an orbit inclined by more than 7$^\circ$ (21$^\circ$) relative to HR 8799 b(c)'s 
orbit: few orbital solutions consistent with the astrometry also place them on coplanar orbits.
Adopting a less restrictive definition for ``acceptably-fitting" orbits does not undo any of these trends, although there are more 
orbit combinations making the planets coplanar.  Adopting the median parameter 
value or MLO instead of the more conservative 68\% confidence interval likewise does not change these results.

These results provide valuable input for constraining the mass of the HR 8799 planetary system.  
Longer-term astrometric monitoring of HR 8799 
\citep[i.e.][]{Konopacky2011} will better clarify the planets' orbital properties.  Limits on the planets' dynamical 
masses will provide crucial input for planet cooling models 
and even more firmly establish HR 8799 as a benchmark system to understand the properties of 
young, self-luminous planets.

Finally, this work and other recent studies of HR 8799 \citep{Soummer2011,Lafreniere2009,Fukagawa2009} 
clearly demonstrate the value of publicly archiving data on advanced telescopes.  In our case, detecting at 
least two HR 8799 planets (HR 8799 bc) was rather straightforward and
 did not require advanced image processing techniques developed well after the data were taken.  
As data for Keck and many other 8--10 m class telescopes are now archived, they provide 
an indispensible resource with which to confirm and characterize directly imaged planets like 
HR 8799's and other substellar companions.
\acknowledgements 
We thank Christian Marois, Scott Kenyon, and the anonymous referee for helpful comments.
This research has made use of the Keck Observatory Archive (KOA), which is operated by the 
W. M. Keck Observatory and the NASA Exoplanet Science Institute (NExScI), under contract with the National Aeronautics and Space Administration. 
We are extremely grateful to the NExScI/KOA staff for developing and maintaining
 the NIRC2 archive.  TC is supported by a NASA Postdoctoral Fellowship; SM is supported by an 
Astronomy Center for Theory and Computation Prize Fellowship at the University of Maryland.

{}

\begin{deluxetable}{lcllccccccc}
\setlength{\tabcolsep}{0.03in}
\tablecolumns{4}
\tablecaption{HR 8799 Planet Photometric/Astrometric Properties}
\tiny
\tablehead{{}&{HR 8799 b}&{HR 8799 c}&{HR 8799 d} \\
{Parameter}&{}&{}&{}}
\tiny
\startdata
\textbf{Measured}\\
\textit{Photometry}\\
$m(H)$ on 2005-07-15 & 18.05$\pm$ 0.09 & 17.06 $\pm$ 0.13 & 16.71 $\pm$ 0.24\\
\textit{Astrometry}\\
2005-07-15 ([E,N])\arcsec{} & 1.496, 0.856 ($\pm$ 0.005) &$-$0.713, 0.630 ($\pm$ 0.005) &$-$0.087, $-$0.578 ($\pm$ 0.010)\\
2010-10-30 ([E,N])\arcsec{} & 1.546, 0.748 ($\pm$ 0.005) &$-$0.598, 0.737 ($\pm$ 0.005) &$-$0.283, $-$0.567 ($\pm$ 0.005)\\
\\
\textbf{Derived}\\
\textit{MLO, med.,[68\% C.I.]}\\
($\chi^{2}_\mathrm{lim}$ $\le$ $\chi^{2}_\mathrm{min}$+1)\\
$a$ (AU), full & 71.0, 109.9 [69.7,164.9] & 37.2, 38.0 [35.5,42.0] & 26.2, 27.3 [24.4,31.5]\\
$\prime\prime$ 4:2:1 resonance  & 68.1, 68.8 [67.5,70.8] & 42.5, 43.2 [42.1,44.4] & 27.3, 27.3 [26.4,28.1]\\
$i$ ($^{\circ}$), full & 14.1, 34.9 [12.3,43.0] & 21.7, 19.9 [12.5,25.8] & 37.1, 37.9 [31.6,41.6]\\
$\prime\prime$ 4:2:1 resonance & 8.5, 9.5 [4.9,14.8] & 25.8, 27.6 [25.2,28.8] & 38.1, 37.9 [36.0,39.1]\\

$e$, full & 0.02, 0.27 [0.02,0.49] & 0.01, 0.08 [0.03,0.13] & 0.04, 0.09 [0.03,0.23]\\
$\prime\prime$ 4:2:1 resonance & 0.01, 0.02 [0.0,0.03] & 0.12, 0.14 [0.11,0.17] & 0.03, 0.04 [0.01,0.08]]\\

$\Omega$ ($^{\circ}$), full & 149.7, 141.0 [40.9,161.4] & 128.8, 122.2 [62.0,152.4] & 56.8, 53.5 [43.3,63.0]\\
$\prime\prime$ 4:2:1 resonance & 163.2, 87.1 [22.9,158.9] & 147.4, 131.9, [104.8,158.5] & 56.9, 54.3 [46.0,60.2]]\\

\\
($\chi^{2}_\mathrm{lim}$ $\le$ $\chi^{2}_\mathrm{min}$+[5,5,3.5])\\ 
$a$ (AU), full & 70.7, 80.8 [68.2,117.2] & 38.2, 39.8 [36.6,46.2] & 26.2, 27.0 [23.0,31.0]\\
$\prime\prime$ 4:2:1 resonance  & 68.0, 68.5 [66.4,71.0] & 42.7, 42.9 [41.3,44.6] & 27.3, 27.1 [26.0,28.2]\\
$i$ ($^{\circ}$), full & 15.2, 24.5 [10.4,36.6] & 19.7, 23.2 [14.0,31.3] & 38.1, 36.9 [27.1,41.2]\\
$\prime\prime$ 4:2:1 resonance & 10.8, 11.5 [6.1,17.1] & 28.8, 27.4 [24.6,30.2] & 37.1, 37.7 [34.6,39.9]\\
$e$, full & 0.01, 0.13 [0.02,0.34] & 0.01, 0.07 [0.02,0.15] & 0.01, 0.15 [0.04,0.32]\\
$\prime\prime$ 4:2:1 resonance & 0.01, 0.03 [0.01,0.06] & 0.02, 0.09 [0.02,0.14] & 0.0, 0.05 [0.01,0.11]\\
$\Omega$ ($^{\circ}$), full & 142.7, 119.5 [32.4,160.5] & 124.7, 86.8 [37.1,142.5] & 52.5, 58.4 [41.0,98.5]\\
$\prime\prime$ 4:2:1 resonance & 142.1, 78.2 [21.9,154.3] & 61.1, 70.5 [42.2,140.5] & 60.9, 55.7 [44.1,67.2]
 \enddata
\tablecomments{\noindent 
\textit{Measured parameters} -- Our photometric uncertainties consider both the signal-to-noise of our detections
 and the absolute flux calibration uncertainties; astrometric uncertainties consider the SNR, astrometric calibration uncertainty 
(e.g. 0.5 pixels in x and y, see Sect. 2), etc.
\textit{Derived parameters} -- The three column entries are \textit{MLO} (the ``most likely orbit" (see \S 3), 
\textit{med.} (the median parameter value), and \textit{[68\% C.I.]} (the 68\% confidence interval).
} 
\label{hr8799prop}
\end{deluxetable}

\end{document}